\documentclass[fleqn,usenatbib]{mnras}

\usepackage{newtxtext,newtxmath,verbatim}

\usepackage[T1]{fontenc}

\DeclareRobustCommand{\VAN}[3]{#2}
\let\VANthebibliography\thebibliography
\def\thebibliography{\DeclareRobustCommand{\VAN}[3]{##3}\VANthebibliography}

\usepackage{graphicx}	
\usepackage{amsmath}	
\usepackage{ulem}
\usepackage[flushleft]{threeparttable}
\usepackage{array} 
\usepackage[english]{babel}
\usepackage{blindtext}
\usepackage{subcaption}
\usepackage{epstopdf}

\newcommand{\ttime}{Tv_{\rm A}/L_{\rm x}}
\newcommand{\estar}{\epsilon^\ast}
\newcommand{\et}{\epsilon_{\rm T}}

\newcommand{\echi}{\epsilon_{\chi}}
\newcommand{\zetaall}{\zeta(\estar,\et)}
\newcommand{\figg}[1]{Fig.~\ref{fig:#1}}

\title[The Role of Electric Dominance for Particle Injection in Relativistic Reconnection]{The Role of Electric Dominance for Particle Injection in Relativistic Reconnection}

\author[S. Gupta, N. Sridhar, {\&} L. Sironi (2024)]{
Sanya Gupta,$^{1}$\thanks{E-mail: sg4038@columbia.edu}
Navin Sridhar,$^{2,3}$\thanks{E-mail: nsridhar@stanford.edu}
Lorenzo Sironi$^{4,5}$\thanks{E-mail: lsironi@astro.columbia.edu}
\\$^{1}$Barnard College, Columbia University, 3009 Broadway, New York, NY 10027, USA
\\$^{2}$Department of Physics, Stanford University, 382 Via Pueblo Mall, Stanford, CA 94305, USA
\\$^{3}$Kavli Institute for Particle Astrophysics \& Cosmology, P.O. Box 2450, Stanford University, Stanford, CA 94305, USA
\\$^{4}$Department of Astronomy and Columbia Astrophysics Laboratory, Columbia University, 550 W 120th St, New York, NY 10027, USA
\\$^{5}$Center for Computational Astrophysics, Flatiron Institute, 162 5th Avenue, New York, NY 10010, USA}

\date{Accepted XXX. Received YYY; in original form ZZZ}

\pubyear{2024}

\begin{document}
\label{firstpage}
\pagerange{\pageref{firstpage}--\pageref{lastpage}}
\maketitle

\begin{abstract}
Magnetic reconnection in relativistic plasmas---where the magnetization $\sigma\gg1$---is regarded as an efficient particle accelerator, capable of explaining the most dramatic astrophysical flares. We employ two-dimensional (2D) particle-in-cell simulations of relativistic pair-plasma reconnection with vanishing guide field and outflow boundaries to quantify the impact of the energy gain occurring in regions of electric dominance ($E>B$) for the early stages of particle acceleration (i.e., the ``injection'' stage). 
We calculate the mean fractional contribution $\zetaall$ by $E>B$ fields to particle energization up to the injection threshold energy, $\estar=\sigma/4$; here, $\et$ is the particle energy at time $T$. We find that $\zeta$ monotonically increases with $\sigma$ and $\et$; for $\sigma\gtrsim 50$ and $\et/\sigma\gtrsim 8$, we find that $\gtrsim 80\%$ of the energy gain obtained before reaching $\estar=\sigma/4$ occurs in $E>B$ regions. We find that $\zeta$ is independent of simulation box size $L_x$, as long as $\et$ is normalized to the maximum particle energy, which scales as $\epsilon_{\rm max}\propto L_{\rm x}^{1/2}$ in 2D. The distribution of energy gains $\echi$ acquired in $E>B$ regions can be modeled as $dN/d\echi\propto\echi^{-0.35}\exp[-(\echi/0.06\,\sigma)^{0.5}]$.
Our results help assess the role of electric dominance in relativistic reconnection with vanishing guide fields, which may be realized in the magnetospheres of black holes and neutron stars.
\end{abstract}




\section{Introduction} \label{introduction}
Magnetic reconnection in relativistic plasmas---where the magnetization $\sigma\gg1$ (defined as the ratio of magnetic enthalpy density to plasma enthalpy density)---is regarded as an efficient particle accelerator, capable of explaining the most dramatic astrophysical flares. Since the very first studies, the question of whether
relativistic reconnection is capable of generating non-thermal particles---and the mechanisms of particle acceleration---has been a crucial focus \citep[for reviews of relativistic reconnection, see] []{hoshino_lyubarsky_12,kagan_15,guo_review_20,guo_review}.

Electric fields in reconnection can be conveniently divided between ideal fields $\boldsymbol{E}_{\rm I}=-(\boldsymbol{v}_{\rm fl}/c) \times \boldsymbol{B}$, where $\boldsymbol{v}_{\rm fl}$ is the fluid velocity, and non-ideal fields $\boldsymbol{E}_{\rm N}=\boldsymbol{E}-\boldsymbol{E}_{\rm I}$, which cannot be captured in ideal magnetohydrodynamics. Non-ideal regions prominently appear at/near X-points, where field lines tear and reconnect.  In fact, X-points have long been considered prime candidates for generating non-thermal spectra in relativistic reconnection \cite[e.g.,][]{larrabee_03,zenitani_01,lyubarsky_liverts_08}.
Non-ideal fields necessarily occur in regions of electric dominance ($E>B$) or in the presence of a field-aligned electric field ($E_\parallel=\boldsymbol{E}\cdot \boldsymbol{B}/B\neq 0$). While such conditions are sufficient, regions where $\boldsymbol{E}\neq \boldsymbol{E}_{\rm I}$ also exist without electric dominance or $E_\parallel$ \citep{Totorica_2023}. 

While intense non-ideal fields are confined to small regions, ideal electric fields driven by fluid motions pervade the reconnection volume and are thus prime candidates for governing particle acceleration to the highest energies \citep{guo_15,guo_19, uzdensky_22}. Fast outflows in the reconnection downstream carry strong ideal fields, with $E \sim B_0$ (here, $B_0$ is the strength of the reconnecting field). Particles can be accelerated by scattering back and forth between coalescing plasmoids via a Fermi-like process \citep{G14,guo_15}, or between plasmoids and fast outflows (see \citealt{Nalewajko_2015} for a comprehensive review of acceleration mechanisms in two-dimensional [2D] reconnection). The converging upstream inflows carry an ideal electric field $E \sim (v_{\rm in}/c) B_0$, where $v_{\rm in}\sim 0.1\,c$ is the inflow velocity. The latter plays an important role for particle acceleration to the highest energies in three-dimensional [3D] relativistic reconnection \citep{zhang_21,zhang_23}.

While it is widely accepted that most of the energy gain of ultra-relativistic particles comes from ideal fields, non-ideal fields have been invoked in the early stage of particle energization. In this phase, acceleration is a rapid, one-shot, ``injection'' process, which boosts the particles from the low, e.g., non-relativistic, upstream energies up to ultra-relativistic energies ($\gamma\sim\sigma\gg1$). 
For moderate/strong guide fields ($B_{\rm g}\gtrsim (v_{\rm in}/c) B_0$), $E_\parallel$ plays a dominant role for particle injection \citep{sironi_22,French_2023}. 
In weak guide fields ($B_{\rm g}\lesssim (v_{\rm in}/c) B_0$),  most of the
particles ending up with high energies have passed
through regions of electric dominance, $E>B$ \citep{zenitani_01,larrabee_03,lyubarsky_liverts_08,sironi_22,chernoglazov_23}. However, it remains debated whether such regions provide most of the energization during the injection stage. \citet{French_2023} found that Fermi reflection and pick-up acceleration in the snapping reconnected field lines dominate over direct $E>B$ acceleration in the injection phase of $B_{\rm g}=0$ reconnection. Still, broadly-defined non-ideal fields ($\boldsymbol{E}\neq\boldsymbol{E}_{\rm I}$, but not necessarily $E>B$) appear to be essential for particle injection even for vanishing guide fields \citep{Totorica_2023}. 

In this work, we employ 2D particle-in-cell (PIC) simulations with outflow boundaries to quantify the impact of the energy gain occurring in regions of electric dominance ($E>B$) for the early stages of particle acceleration in relativistic pair-plasma reconnection with vanishing guide field ($B_{\rm g}=0$). We calculate the mean fractional contribution $\zetaall$ by $E>B$ fields to particle energization up to a characteristic injection threshold $\estar=\sigma/4$; here, $\et$ is the particle energy at time $T$. We find that $\zeta$ monotonically increases with $\sigma$ and $\et$; for $\sigma\gtrsim 50$ and $\et/\sigma\gtrsim 8$, we find that $\gtrsim 80\%$ of the energy gain obtained before reaching $\estar=\sigma/4$ occurs in $E>B$ regions. We find that $\zeta$ is independent of box size $L_x$, as long as $\et$ is normalized to the maximum particle energy, which scales as $\epsilon_{\rm max}\propto L_{\rm x}^{1/2}$ in 2D. The distribution of energy gains $\echi$ acquired in $E>B$ regions can be modeled as $dN/d\echi\propto\echi^{-0.35}\exp[-(\echi/0.06\,\sigma)^{0.5}]$.

This paper is organized as follows. In Section \ref{simulation} we describe the simulation setup. In Section \ref{results} we present our results, discussing specifically the dependence on magnetization and box size. Finally, our conclusions, the astrophysical implications of our work, and future steps are outlined in Section \ref{conclusions}.

\section{PIC Simulation Setup} \label{simulation}
Our simulations are performed with the particle-in-cell (PIC) code \texttt{TRISTAN-MP} \citep{S05} and we use a Vay pusher \citep{V08} to advance the particle momenta. We employ a 2D simulation domain in the $x-y$ plane, but we track all components of velocities, electric currents and electromagnetic fields.
We initialize the system with a cold electron-positron plasma with rest-frame density of $n_{0} = 64$ particles per cell (including both species). We resolve the plasma inertial length/skin depth ($c/\omega_{\rm p}$, where $\omega_{\rm p} = \sqrt{4\pi n_{0} e^{2}/m}$ is the plasma frequency, and $m$ and $e$ are the positron mass and charge) with 5 cells. The timestep is $0.09\,\omega_{\rm p}^{-1}$. In Appendix \ref{spatial_res}, we show convergence of our results with respect to spatial resolution and $n_0$.

The magnetic field is initialized in force-free equilibrium: the field has strength $B_{0}$ and its direction rotates from $+\hat{x}$ to $-\hat{x}$ across a current sheet located at $y=0$. The $x$-component of the field varies as $B_x=-B_0 \tanh(y/\Delta)$, where the sheet thickness is $\Delta \simeq 11 \,c/\omega_{\rm p}$. The field strength $B_{0}$ is parameterized by the magnetization
\begin{equation}
    \sigma = \frac{B^{2}_{0}}{4\pi n_{0}mc^{2}} = \left(\frac{\omega_{c}}{\omega_{\rm p}}\right)^{2}
\end{equation} 
where $\omega_{c} = eB_{0}/mc$ is gyro-frequency. We vary the magnetization, exploring $\sigma = 12.5, 50, 200$, with $\sigma = 50$ being our reference case.

We trigger reconnection in the middle of the initial current sheet \citep{sironi_16}, and evolve the system until $T \sim 5.1\,L_{\rm x}/v_{\rm A}$, where the Alfv\'en speed $v_{\rm A}=c\sqrt{\sigma/(1+\sigma)}$. The fiducial box half-length along the $x$-direction of reconnection outflows is $L_{\rm x}/(c/\omega_{\rm p}) = 768$. We also perform larger simulations with $L_{\rm x}/(c/\omega_{\rm p}) = 1536$ and smaller runs with $L_{\rm x}/(c/\omega_{\rm p}) = 384$ to assess the dependence on the domain size.
Along the $y$-direction of reconnection inflows, we employ two moving injectors---that constantly introduce fresh magnetized plasma into the simulation domain---and an expanding simulation box  \citep{sironi_16}. In most of our simulations we adopt open boundaries for fields and particles along the $x$-direction of reconnection outflows \citep{sironi_16}. To facilitate comparison with earlier papers, we also include three simulations (one for each magnetization) having periodic boundaries along $x$. There, we do not perturb the initial current sheet to trigger reconnection---rather, we start with a thinner sheet so reconnection develops spontaneously. Our numerical and physical parameters are listed in Table \ref{tab:input_parameters}.

\begin{table}
 \caption{Table of numerical and physical parameters}
 \begin{center}
 \begin{tabular}{cccc}
		\hline
        \hline
		\relax Boundary type$^{[1]}$ & ${\sigma}^{[2]}$ & ${L_{\rm x}/(c/\omega_{\rm p})}^{[3]}$ & ${n_0}^{[4]}$\\
		\hline
        \hline 
		outflow & 12.5 & 384 & 64\\
		outflow & 50 & 384 & 64\\
		outflow & 200 & 384 & 64\\
        \hline 
		outflow & 12.5 & 768 & 16\\
		outflow & 50 & 768 & 16\\
		outflow & 200 & 768 & 16\\
        \hline
		outflow & 12.5 & 768 & 64\\
        outflow & 50 & 768 & 64\\
        outflow & 200 & 768 & 64\\
        \hline
        outflow & 12.5 & 768 & 90\\
		outflow & 50 & 768 & 90\\
        outflow & 200 & 768 & 90\\
        \hline
		periodic & 12.5 & 768 & 64\\
        periodic & 50 & 768 & 64\\
        periodic & 200 & 768 & 64\\
		\hline
        outflow & 12.5 & 1536 & 64\\
        outflow & 50 & 1536 & 64\\
        outflow & 50 & 1536 & 64\\
		\hline
	\end{tabular}
 \end{center}
 \label{tab:input_parameters}
 \begin{tablenotes}
 \small
 \item \relax \textit{Note}: All simulations are performed for the same duration of $T \sim 5.1\,L_{\rm x}/v_{\rm A}$, with the same spatial resolution of 5 cells per $c/\omega_{\rm p}$ and zero guide field. The description of each column is as follows: ${^{[1]}}$ type of boundary; ${^{[2]}}$ upstream magnetization; ${^{[3]}}$ half-length of the computational domain along $x$, in units of $c/\omega_{\rm p}$; ${^{[4]}}$ initial particle number density in the upstream.
 \end{tablenotes}
\end{table} 


\begin{figure}
\includegraphics[width=0.5\textwidth]{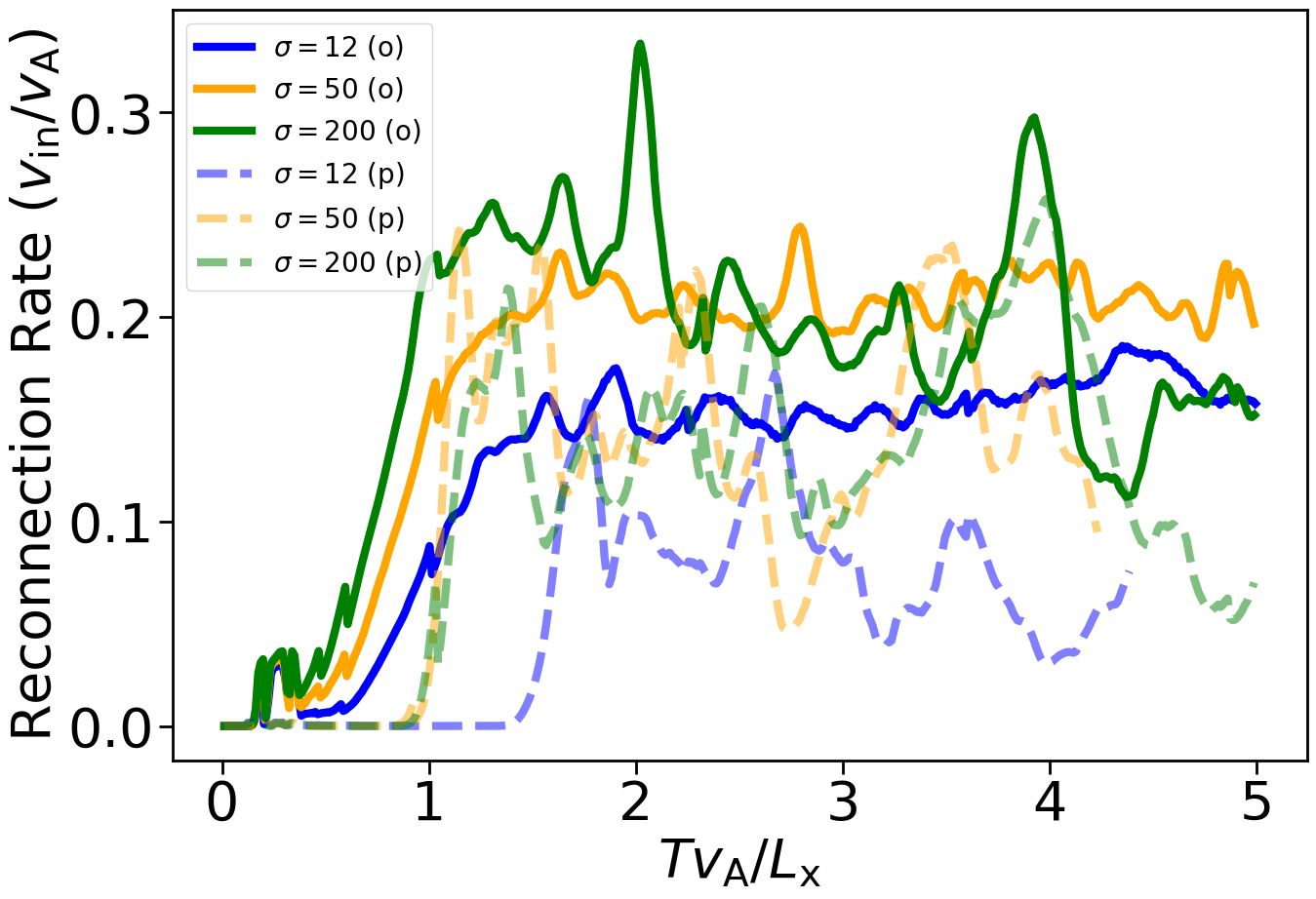}
 \caption{Reconnection rate (i.e., inflow velocity) in units of the Alfv\'{e}n speed, $v_{\rm in}/v_{\rm A}$, as a function of time (in units of $L_{\rm x}/v_{\rm A}$). Colors represent different magnetizations: $\sigma$ = 12.5 (blue),  50 (orange), 200 (green). We show cases with outflow boundary conditions (denoted `o') using solid curves and cases with periodic boundaries (denoted `p') using dashed translucent curves.}
 \label{fig:rec_rate}
\end{figure}

\begin{figure*}
  \includegraphics[width=0.8\textwidth]{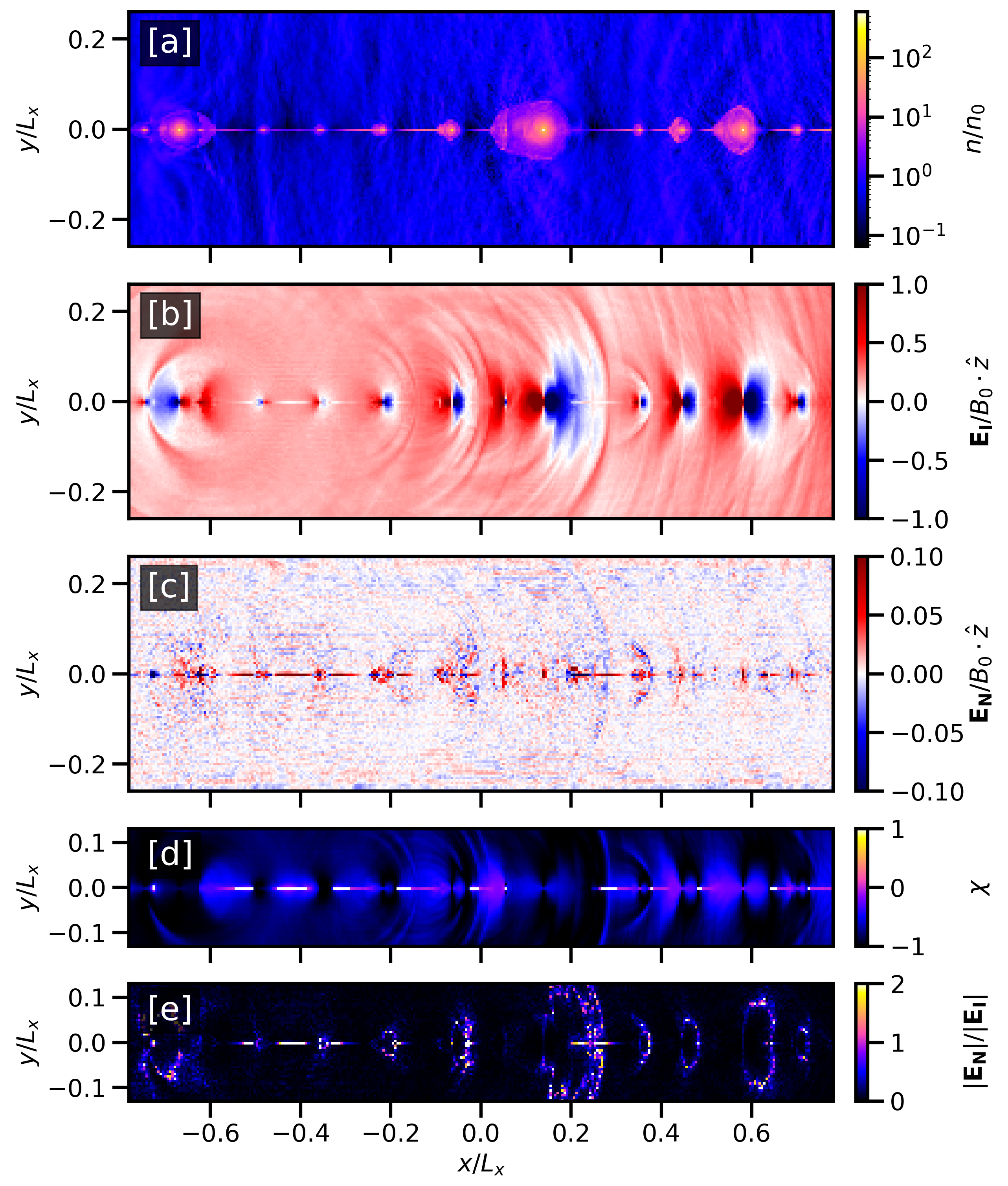}
  \caption{Structure of the reconnection layer at time $Tv_{\rm A}/L_{\rm x} \sim 2.8$ (after  the system has settled into a quasi-steady state) for our fiducial outflow run with magnetization $\sigma = 50$ and box size $L_{\rm x}=768\,c/\omega_{\rm p}$. We show a subset of the domain: the range in $x$ is  $-0.8 < x/L_{\rm x} < 0.8$, while along $y$ we focus on $-0.25 < y/L_{\rm x} < 0.25$ for panels [a],[b],[c] and $-0.15 < y/L_{\rm x} < 0.15$ for panels [d],[e].
We present: [a] the particle number density in units of the upstream rest-frame density $n_{0}$; [b] the $z$-component of the ideal electric field, normalized to $B_0$; [c] the $z$-component of the non-ideal electric field, normalized to $B_0$; [d] the parameter $\chi = (\boldsymbol{E}^2-\boldsymbol{B}^2)/(\boldsymbol{E}^2+\boldsymbol{B}^2)$, where regions of electric dominance ($\chi > 0$) are colored in white; [e] the ratio of non-ideal to ideal electric fields, where regions having $\lvert \boldsymbol{E}_{\rm N} \rvert > 2\, \lvert \boldsymbol{E}_{\rm I} \rvert$ are colored in white.
}
 \label{fig:2d_img}
\end{figure*}

\section{Results} \label{results}

\subsection{Reconnection Rate and Structure of the Layer} \label{2d_visual}
In Fig.~\ref{fig:rec_rate}, we show the reconnection rate as a function of time for different magnetizations and types of boundary conditions. We define the reconnection rate as the upstream plasma’s inflow velocity $v_{\rm in}$ into the layer. This  is computed by taking the spatial average of the inflow velocity $v_{\rm y}$ over a rectangular box excluding 100 cells at each $x$-boundary ($-0.87 \leq x/L_{\rm x} \leq 0.87$) and ranging over $-0.20 \leq y/L_{\rm x} \leq -0.15$.\footnote{We exclude 100 cells at each $x$-boundary since this is the region where the outflow boundary conditions are enforced.} The inflow velocity $v_{\rm in}$ is given in units of the Alfv\'en speed, $v_{\rm A} =c \sqrt{\sigma/(\sigma+1)}$. In the case of outflow boundary conditions (solid lines), the reconnection rate settles into a quasi-steady value at $\ttime \gtrsim 1.5$; thereafter, we shall say that the system has achieved a quasi-steady state. In contrast, cases with periodic boundaries do not attain a steady state. There, plasma and magnetic flux accumulate in plasmoids that grow bigger and bigger as the simulation progresses; at some point, the large plasmoids inhibit further inflow of plasma into the layer and the reconnection rate drops, as indeed displayed by the dashed translucent curves.


Fig.~\ref{fig:2d_img} shows the structure of the reconnection layer in our reference simulation ($\sigma = 50$, $L_{\rm x} = 768\, c/\omega_{\rm p}$) at $Tv_{\rm A}/L_{\rm x} \sim 2.8$, when the system has settled into a quasi-steady state. We show a subset of the domain: the range in $x$ is  $-0.8 < x/L_{\rm x} < 0.8$, while along $y$ we focus on $-0.25 < y/L_{\rm x} < 0.25$ for panels [a],[b],[c] and $-0.15 < y/L_{\rm x} < 0.15$ for panels [d],[e].
Panel [a] presents the plasma density, showing evidence of reconnection plasmoids. At this time, three large plasmoids are visible at $x/L_{\rm x} \sim -0.7,0.1,0.6$, with smaller plasmoids scattered all along the reconnection layer. 

The rest of the figure focuses on the electromagnetic field properties. The electric field is divided into two components: the ideal field $\boldsymbol{E}_{\rm I}=-(\boldsymbol{v}_{\rm fl}/c) \times \boldsymbol{B}$, where $\boldsymbol{v}_{\rm fl}$ is the fluid velocity, and the non-ideal field $\boldsymbol{E}_{\rm N}=\boldsymbol{E}-\boldsymbol{E}_{\rm I}$. Here, $\boldsymbol{v}_{\rm fl}$ is the local mean of the particle velocities, obtained for each cell by averaging over the particles in the neighboring $5\times 5$ cells. The $z$-component of the ideal field is shown in panel [b], whereas the $z$-component of the non-ideal field in panel [c]; both are normalized to the initial magnetic field $B_{0}$. The typical magnitude of the ideal field in the upstream region is $\sim 0.1\,B_{0}$. The ideal field is stronger in reconnection plasmoids, where it shows a characteristic signature: it is negative on the leading side (i.e., the direction of plasmoid motion) and positive on the trailing side. The peak value of the non-ideal field is $\sim 0.1\,B_{0}$, and it is  concentrated in the thinnest regions of the current sheet.

In panel [d] of Fig.~\ref{fig:2d_img} we plot $\chi = (\boldsymbol{E}^2-\boldsymbol{B}^2)/(\boldsymbol{B}^2+\boldsymbol{E}^2)$ and we color all regions having $\chi > 0$ (i.e., regions of electric dominance) in white.\footnote{We note that \citet{sironi_22} defined $\chi = (\boldsymbol{B}^2-\boldsymbol{E}^2)/(\boldsymbol{B}^2+\boldsymbol{E}^2) $, which is opposite in sign to the choice we adopt here.} Finally, we show the ratio $\lvert \boldsymbol{E}_{\rm N} \rvert / \lvert \boldsymbol{E}_{\rm I} \rvert$ in panel [e]; there, we color in white the regions where  $\lvert \boldsymbol{E}_{\rm N} \rvert / \lvert \boldsymbol{E}_{\rm I} \rvert> 2$. By comparing panels [d] and [e], one sees that a number of regions of electric dominance ($\chi>0$) also host strong non-ideal fields ($\lvert \boldsymbol{E}_{\rm N} \rvert / \lvert \boldsymbol{E}_{\rm I} \rvert> 2$). While the overlap is not exact, one might argue that electric dominance appears to be a good proxy for identifying regions with strong non-ideal fields (more precisely, where non-ideal fields greatly dominate over ideal ones), near the $y=0$ reconnection plane.

\subsection{Assessment of Electric Dominance} \label{contribution}

In this subsection, we quantify the fractional contribution of $E>B$ regions to particle energization. We define regions of electric dominance as $\chi > 0$. For each particle, we define $\epsilon_{\rm tot} = \gamma - 1$ as its total kinetic energy (in units of the particle rest mass; hereafter, simply ``energy'') and $\epsilon_{\rm \chi} = \gamma_{\rm \chi} - 1$ as the amount of kinetic energy acquired in regions of electric dominance; equivalently, as the work done by $E>B$ electric fields. We then build a statistical assessment of the role of electric dominance as follows (a similar strategy was employed by \citet{Totorica_2023}, to assess the role of non-ideal fields): at time $T$, we bin particles as a function of their energy, $\et$; for the particles in each bin, we track their evolution up to $T$ and detect the time when their energy reached a pre-determined threshold, which we shall define as $\epsilon^\ast$ and call ``injection threshold''; at the time when  $\epsilon_{\rm tot}=\estar$, we compute for each particle the ratio $\epsilon_{\chi} /\epsilon_{\rm tot} $, i.e., the fractional contribution of $\chi>0$ regions to particle energization up to the injection threshold $\estar$; we finally compute the average of this ratio over the particles in the same $\et$ bin, which we define as
\begin{equation} \label{eq:e_avg}  
\zeta(\estar,\et)\equiv\overline{(\epsilon_{\rm \chi}/\epsilon_{\rm tot})\mid}_{\epsilon_{\rm tot}=\epsilon^\ast}~.
\end{equation}
In order to minimize statistical fluctuations, rather than showing results obtained from the analysis of a single snapshot at time $T$, we repeat this procedure for a few different snapshots (separated by $90\,\omega_{\rm p}^{-1}$) and then average their outcomes. The time interval over which we average is indicated in the figures below.

For simulations with outflow boundaries, the results presented below exclude particles residing less than 100 cells from the two $x$-boundaries, i.e., where we impose the outflow boundary conditions. For cases with periodic boundaries, we include the whole $x$-extent of the domain. Most of the results in this subsection refer to outflow simulations, while we present in Appendix \ref{periodic} a more detailed analysis of periodic cases, for comparison with \citealt{Totorica_2023}. 

We point out that our calculation of $\zeta(\estar,\et)$ does not differentiate on the basis of the particle lifetime (i.e., how long a given particle needed to reach $\estar$ or $\et$, following its first interaction with the layer). In Appendix \ref{age}, we show that  $\zeta(\estar,\et)$ is the same for particles having a wide range of lifetimes.

\begin{figure*}
 \includegraphics[width=\textwidth]{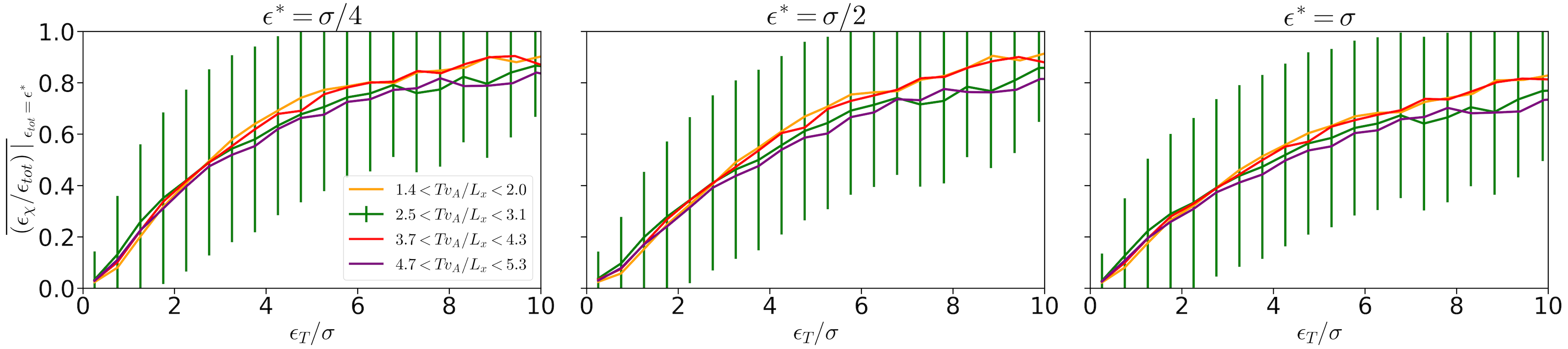}
 \caption{Average fractional contribution of $E>B$ regions to particle energization, for our fiducial simulation ($\sigma=50$, $L_{\rm x}=768\,c/\omega_{\rm p}$, and outflow boundaries). We plot  $\zeta(\estar,\et)\equiv\overline{(\epsilon_{\rm \chi}/\epsilon_{\rm tot})\mid}_{\epsilon_{\rm tot}=\epsilon^\ast}$ on the vertical axis and $\et/\sigma$ on the horizontal axis. Each panel refers to a different injection threshold, $\epsilon^\ast = \sigma/4, \sigma/2, \sigma$ from left to right. In each panel, different curves refer to different time intervals over which $\zeta(\estar, \et)$ is averaged: $1.4 < Tv_{\rm A}/L_{\rm x} < 2.0, 2.5 < Tv_{\rm A}/L_{\rm x} < 3.1, 3.7 < Tv_{\rm A}/L_{\rm x} < 4.3, 4.7 < Tv_{\rm A}/L_{\rm x} < 5.3$, respectively in orange, green, red, and purple. Vertical green lines show the standard deviation of our measurements in the representative time range $2.5 < Tv_{\rm A}/L_{\rm x} < 3.1$.
 }
 \label{fig:time_evol}
\end{figure*}


\begin{figure*}
 \includegraphics[width=\textwidth]{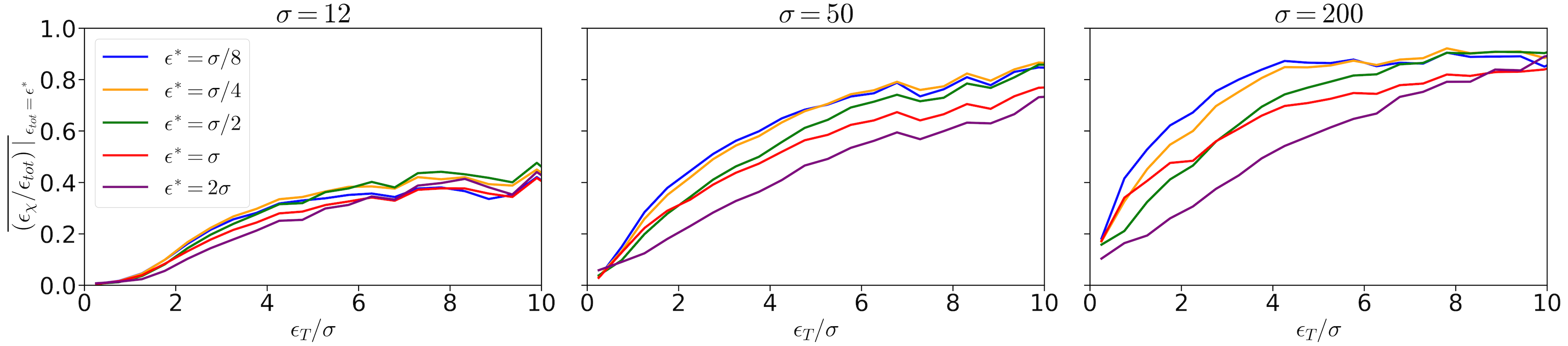}
 \caption{Same as Fig.~\ref{fig:time_evol}, but we vary the magnetization and the injection threshold energy (as indicated by the legend). We use outflow simulations with $L_{\rm x}=768\,c/\omega_{\rm p}$ and consider the fiducial time range $2.5 < Tv_{\rm A}/L_{\rm x} < 3.1$.
 }
 \label{fig:threshold}
\end{figure*}

\begin{figure*}
 \includegraphics[width=\textwidth]{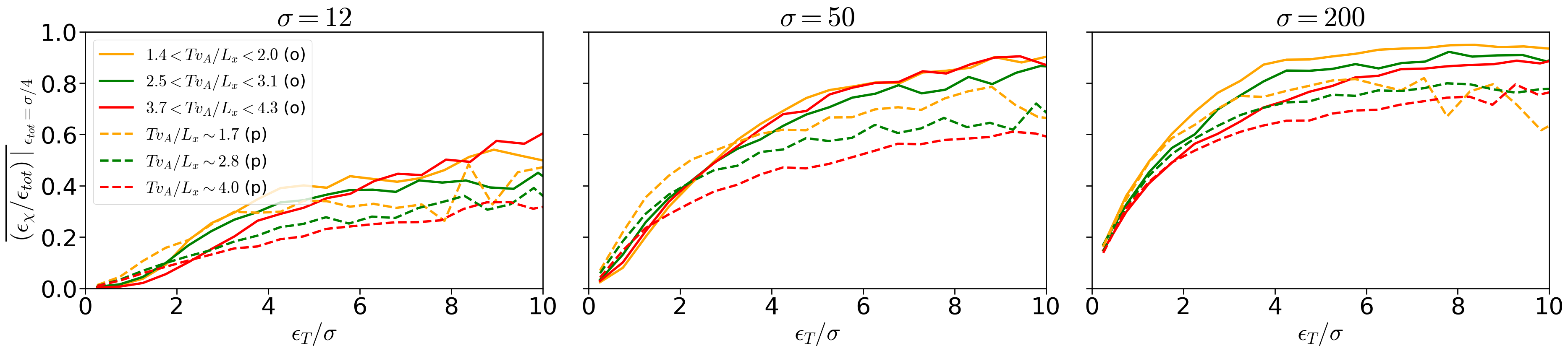}
 \caption{Same as Fig.~\ref{fig:time_evol}, but we vary the type of boundary conditions (solid lines for outflow, dashed lines for periodic) for each magnetization. We use simulations with $L_{\rm x}=768\,c/\omega_{\rm p}$ and consider a few time ranges, as indicated in the legend. For periodic runs, we do not average in time; rather, we consider a snapshot in the middle of the time range used for corresponding outflow cases. We fix the injection threshold at $\estar=\sigma/4$.
}
 \label{fig:time_evol_boundary}
\end{figure*}

\begin{figure*}
 \includegraphics[width=\textwidth]{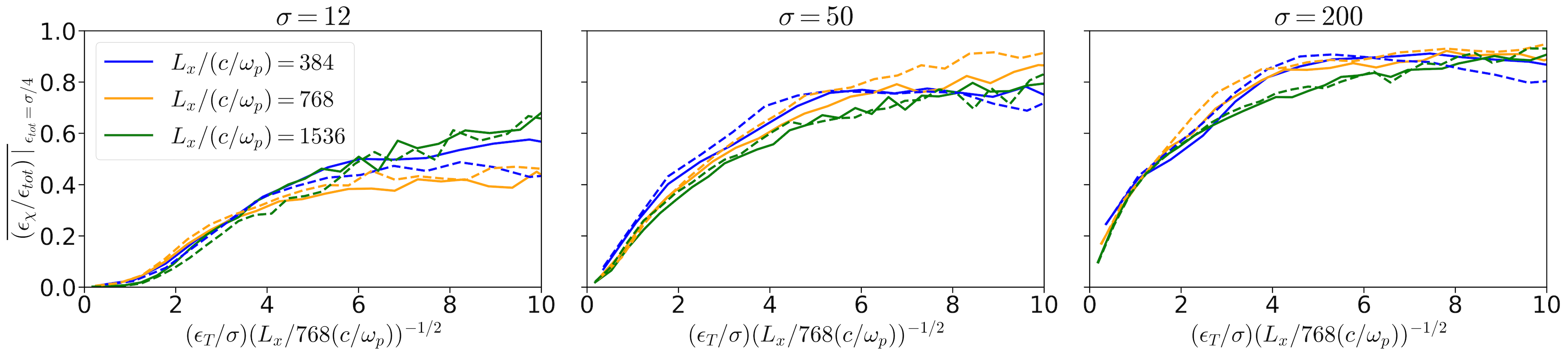}
 \caption{Same as Fig.~\ref{fig:time_evol}, but we vary the simulation box size for each magnetization, as indicated in the legend. We use outflow simulations and consider two time ranges: dashed lines for $1.9 < Tv_{\rm A}/L_{\rm x} < 2.5$, solid lines for $2.5 < Tv_{\rm A}/L_{\rm x} < 3.1$. 
 We fix the injection threshold at $\estar=\sigma/4$. The energy $\et$ on the horizontal axis is normalized to $\sigma L_{\rm x}^{1/2}$, as explained in the text. 
 }
 \label{fig:box_size}
\end{figure*}

Our results are presented in Figs.~\ref{fig:time_evol}-\ref{fig:box_size}. All figures have a similar layout: we plot $\zeta(\estar,\et)$ as a function of $\et/\sigma$. We vary the temporal range of our calculations (Fig.~\ref{fig:time_evol}), the magnetization and the injection energy threshold (Fig.~\ref{fig:threshold}), the type of boundary conditions (Fig.~\ref{fig:time_evol_boundary}), and the domain size (Fig.~\ref{fig:box_size}).

In  Fig.~\ref{fig:time_evol} we focus on our fiducial case, having $\sigma=50$, $L_{\rm x}=768\,c/\omega_{\rm p}$, and outflow boundaries. Each panel refers to a different injection threshold, $\epsilon^\ast = \sigma/4, \sigma/2, \sigma$ from left to right. In each panel, different curves refer to different time intervals over which we average $\zeta(\estar,\et)$: $1.4 < Tv_{\rm A}/L_{\rm x} < 2.0, 2.5 < Tv_{\rm A}/L_{\rm x} < 3.1, 3.7 < Tv_{\rm A}/L_{\rm x} < 4.3, 4.7 < Tv_{\rm A}/L_{\rm x} < 5.3$, respectively in orange, green, red, and purple. 
The figure allows to draw some important conclusions. First, the assessment of the role of electric dominance, as quantified by $\zeta(\estar,\et)$, is remarkably robust to the time range we consider for our calculation, as long as the system has achieved a quasi-steady state ($\ttime\gtrsim 1.5$). Vertical green lines show the standard deviation of our measurements in the representative time range $2.5 < Tv_{\rm A}/L_{\rm x} < 3.1$, which we later adopt as our fiducial time range. The variations of $\zetaall$ over time (different colors in Fig.~\ref{fig:time_evol}) are much smaller than the standard deviation measured at a given time (green error bars). 

Second, $\zetaall$ is an increasing function of $\et/\sigma$, i.e., particles ending up with higher energies at time $T$ have experienced a more prominent role of $\chi>0$ regions in their early energization history (i.e., before crossing the injection threshold $\estar$). For $\estar=\sigma/4$, which we shall adopt as our reference case below, $\zeta\simeq 0.4$ for $\et/\sigma\simeq 2$ and it increases up to $\zeta\simeq 0.8$ for $\et/\sigma\gtrsim 8$. In other words, for particles with $\et/\sigma\gtrsim 8$, a fraction $\sim 80\%$ of the energy they gained before crossing the $\estar=\sigma/4$ injection threshold has been acquired in regions of electric dominance ($\chi>0$).

Third, at fixed $\et/\sigma$, the function $\zeta$ decreases for increasing $\estar$ (from left to right panels in Fig.~\ref{fig:time_evol}). This is expected: the voltage of a given $\chi>0$ region is fixed, and so will the energy gain of a particle interacting with that region; for a higher injection threshold $\estar$, the fractional contribution of electric dominance will necessarily be smaller. This is further displayed in Fig.~\ref{fig:threshold}, where we analyze the dependence of our results on magnetization (from left to right panel, $\sigma=12.5$, 50, 200) and threshold energy (different colors in a given panel, see legend). For each magnetization, the contribution of $\chi>0$ regions systematically drops for increasing $\estar\gtrsim \sigma/4$. This trend is similar to Fig.~2b of \citet{Totorica_2023}. On the other hand, the curves corresponding to $\estar=\sigma/8$ (blue) and $\estar=\sigma/4$ (orange) nearly overlap for all magnetizations. In the following, we will adopt $\estar=\sigma/4$ as our fiducial injection threshold, the same choice as in \citet{Totorica_2023}.
It may appear surprising that bins with $\et<\estar$ are not empty. This is particularly clear for cases with high threshold, e.g., $\epsilon^\ast = 2\sigma$ (purple). Such bins are populated by the few particles whose energy, after exceeding the injection threshold, decreases down to $\et<\estar$.


Fig.~\ref{fig:threshold} also demonstrates that electric dominance plays a greater role for higher magnetizations, as emphasized by \citet{sironi_22}. This holds regardless of the chosen $\estar$. For our fiducial threshold $\estar=\sigma/4$ and a representative energy of $\et/\sigma\simeq 2$, we find that $\zeta\simeq 0.1$ for $\sigma=12.5$, $\zeta\simeq 0.4$ for $\sigma=50$, and $\zeta\simeq 0.6$ for $\sigma=200$. For $\sigma=200$ and $\estar=\sigma/4$, the contribution of $\chi>0$ regions reaches $\zeta\simeq 0.9$ for $\et/\sigma\gtrsim 4$, suggesting that electric dominance plays a dominant role in the injection stage of particles ending up with the highest energies. 

In order to compare our results---mostly based on simulations with outflow boundaries---with earlier papers, which employed periodic boundary conditions \citep{guo_19,sironi_22,French_2023,Totorica_2023}, in Fig.~\ref{fig:time_evol_boundary} we present the outcomes of a set of simulations, having the same numerical and physical parameters, but adopting different (outflow vs. periodic) boundary conditions. We present results for three magnetizations (left to right panels). Outflow cases are shown with solid lines, periodic cases with dashed lines. Different colors refer to different times: outflow results are averaged over an interval of  $\sim 0.6\,L_{\rm x}/v_{\rm A}$, while periodic cases are measured at a snapshot in the middle of the time range used for the corresponding outflow cases.\footnote{We do not need to average in time for the periodic cases because particles keep accumulating in the layer, and the temporal fluctuations within a time interval of $\lesssim 0.5 \,L_{\rm x}/v_{\rm A}$ are minimal.} We find that, as already anticipated in Fig.~\ref{fig:time_evol}, in outflow cases the role of electric dominance, as quantified by $\zetaall$, is generally steady over time. This is particularly clear for $\sigma=12.5$ and $\sigma=50$. For $\sigma=200$, the value of $\zetaall$ appears to slightly decrease  over time, likely a consequence of the fact that the reconnection rate (and so, the reconnection electric field) moderately drops at later times (see solid green line in Fig.~\ref{fig:rec_rate}).
In the periodic cases, the value of $\zetaall$ at early times ($\ttime\sim 1.7$, yellow dashed lines) is comparable to the corresponding outflow runs. However,  $\zetaall$ then 
decreases with time, in parallel to the drop in reconnection rate (Fig.~\ref{fig:rec_rate}). 
It follows that the role of electric dominance can be more reliably assessed in outflow simulations, rather than the periodic cases employed in earlier papers.
Still,  we present in Appendix \ref{periodic} a more detailed analysis of periodic cases, to facilitate comparison with \citealt{Totorica_2023}. There, we show that a dominant fraction of
the energy gain provided by non-ideal fields should be ascribed to regions of electric dominance.

We conclude this section by presenting the dependence of $\zetaall$ on the size of the simulation domain, for a range of simulations (of varying magnetization) with outflow boundary conditions.
We present results, in Fig.~\ref{fig:box_size}, averaged over the fiducial time range $2.5 < Tv_{\rm A}/L_{\rm x} < 3.1$ (solid) and over an earlier time range $1.9 < Tv_{\rm A}/L_{\rm x} < 2.5$ (dashed), to check the robustness of our conclusions. We find that, if the horizontal axis were to be kept at $\et/\sigma$ regardless of box size, the fractional contribution of $\chi>0$ regions would systematically decrease with increasing box size. Instead, we choose to scale the horizontal axis to the expected maximum energy of accelerated particles. In 2D, the highest-energy particles are trapped in plasmoids \citep{werner_16,uzdensky_22}, and their energy increases as the square root of time, due to magnetic moment conservation, coupled with a linear increase in the local field strength as plasmoid cores compress during subsequent plasmoid growth \citep{petropoulou_18,hakobyan_21}. In outflow simulations, the time available for acceleration is comparable to the plasmoid advection time out of the box, $\sim L_{\rm x}/v_{\rm A}$, which implies that the maximum particle energy scales (at fixed $\sigma$) as $\epsilon_{\rm max}\propto L_{\rm x}^{1/2}$. When the energy $\et$ on the horizontal axis is scaled with $\epsilon_{\rm max}$, as in Fig.~\ref{fig:box_size}, we find that the curves of $\zetaall$ nearly overlap, when varying box size at fixed magnetization. In other words, for each magnetization, the fractional contribution of regions of electric dominance only depends on $\et/\epsilon_{\rm max}$. The main exception is the high-energy end of the $\sigma=12.5$ case, where the different colored curves deviate from each other; we note, however, that the trend is not systematic, since the largest box (green) lies above all others, while the smallest one (blue) is in between the intermediate (orange) and the large box.

\subsection{Particle Energy Spectra} \label{spectra}

\begin{figure*}
 \includegraphics[width=\textwidth]{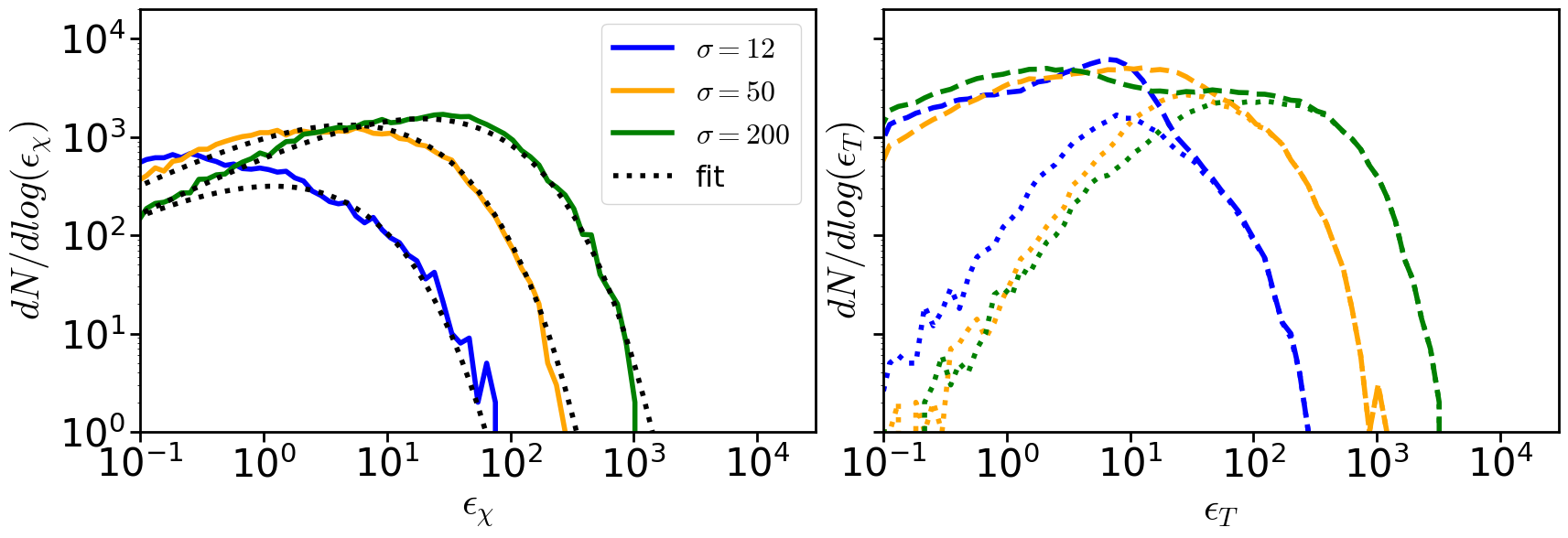}
 \caption{Left panel: the distribution of $\epsilon_{\rm \chi}=\gamma_\chi-1>0$, i.e., of the energy gains obtained in regions of electric dominance before time $T=2.8\,L_{\rm x}/v_{\rm A}$. Right panel: distribution of particle energies for all the particles ($\epsilon_{\rm T}$; solid lines) and for the particles contributing to the left panel ($\epsilon_{\rm T}\mid_{\epsilon_{\rm \chi} > 0}$; dotted lines), measured at $T=2.8\,L_{\rm x}/v_{\rm A}$. Curves are color-coded by the magnetization (see legend). The dotted black lines in the left panel show the best-fit curves using Eq.~\ref{eq:fit}.
 }
 \label{fig:wpar_gam}
\end{figure*}

\begin{figure}
    \centering
    \includegraphics[width=0.5\textwidth]{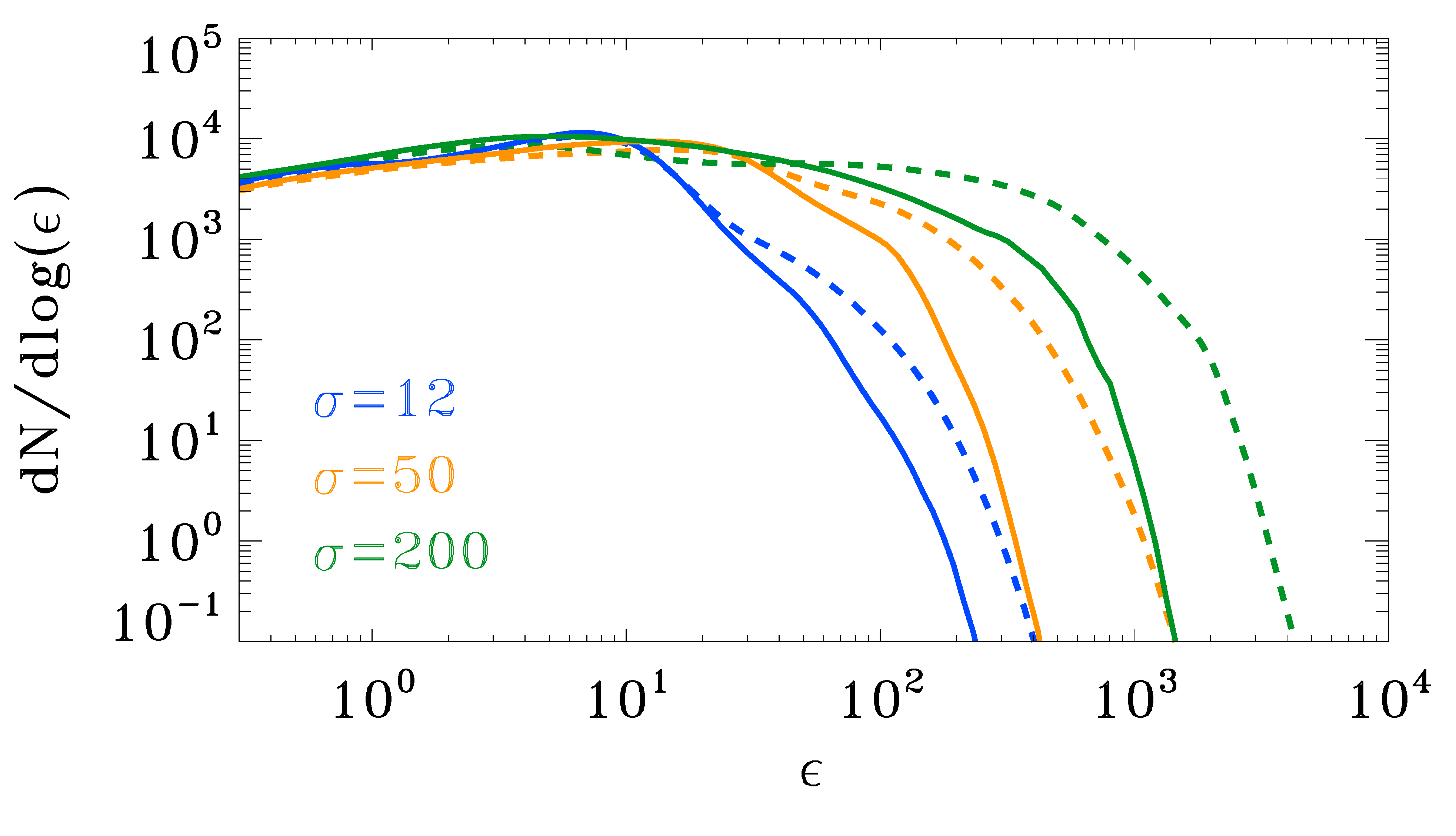}
    \caption{Dashed lines: self-consistent particle spectra from our simulations---time-averaged from $\ttime\simeq 2.0$ to 2.8---as a function of magnetization (see legend). Solid lines: energy spectra of test-particles initialized and evolved as regular particles, but such that their energy is artificially kept fixed while in $E>B$ regions.}
    \label{fig:lorenzo}
\end{figure}

We complement our investigation with an analysis of the particle energy spectra. 
Fig.~\ref{fig:wpar_gam} shows in the left panel the distribution of $\epsilon_{\rm \chi}=\gamma_\chi-1>0$, i.e., of the energy gained in regions of electric dominance before time $T$ (each particle counts at most once, even if it were to interact with $\chi>0$ regions more than one time). We point out that this is not equal to the spectrum of particles residing in $\chi>0$ regions at the time $T$. 
The right panel shows the distribution of particle energies for all the particles ($\epsilon_{\rm T}$; solid lines) and for the particles contributing to the left panel ($\epsilon_{\rm T}\mid_{\epsilon_{\rm \chi} > 0}$; dotted lines). The spectra in both panels refer to positrons, but we have checked that electron spectra are identical, for each magnetization. We have selected $Tv_{\rm A}/L_{\rm x} \sim 2.8$ as a representative time; we extend this analysis to several other times in Appendix \ref{time_energy}. 


The right panel in Fig.~\ref{fig:wpar_gam} confirms that the high-energy end of the particle spectrum is populated by particles that have interacted with $\chi>0$ regions of electric dominance (i.e., dotted and solid lines overlap at high energies), as discussed by \citet{sironi_22}. The left panel illustrates the statistics of the energy gains acquired in $\chi>0$ regions. We find that the curves in the left panel, corresponding to different $\sigma$, can be approximated by a distribution of the form
\begin{equation}
\frac{dN}{d \epsilon_\chi} \propto \epsilon_{\rm \chi}^{B} \exp\left[{-\left(\frac{\epsilon_{\rm \chi}}{A(\sigma)}\right)^{D}}\right]
\label{eq:fit}
\end{equation}
where the best-fit parameters are $B=-0.35$, $D=0.5$ and $A(\sigma)=0.06\,\sigma$ (the fit is based on the two cases of higher magnetization, $\sigma=50$ and $\sigma=200$). It follows that the mean energy gain while in $\chi>0$ regions is $\langle\echi\rangle\simeq 0.18 \,\sigma$ (considering only particles having $\echi>0$), which is consistent with the analysis presented in the previous subsection, namely that $\estar=\sigma/4$ is an appropriate choice for the injection threshold.  

We conclude this section by demonstrating that energization in $\chi>0$ regions plays an important role in shaping the high-energy end of the particle spectrum. The self-consistent particle spectra from our simulations---time-averaged from $\ttime\simeq 2.0$ to 2.8----are shown with dashed lines in \figg{lorenzo}, for different magnetizations (see legend). Solid lines show instead the spectra of ``test-particles''--- not contributing to the electric currents in the simulation, but otherwise initialized and evolved as regular particles. When test-particles pass through $\chi>0$ regions, we artificially constrain their energy to remain unchanged. We do this by updating their momentum in the same way as for regular particles, but then we re-set the magnitude of the momentum vector to be the one prior to the update.\footnote{Note that \citet{sironi_22} adopted a more aggressive strategy, by artificially fixing the Lorentz factor of test-particles in $E>B$ regions at $\gamma-1\sim {\rm few}$. In addition to inhibiting energization in $E>B$ regions, this also artificially reduced the Lorentz factor of the few high-energy test-particles happening to pass through $E>B$ sites, even in case most of their energy was acquired in regions of magnetic dominance \citep{guo_23b}.}
We argue that this is the most suitable way (as compared to, e.g.,  setting $\boldsymbol{E}=0$ in $\chi>0$ regions) to largely preserve the dynamics of test-particles, and yet inhibit their energization while in $E>B$ regions. The spectra of these test-particles (solid lines) are much steeper  than the ones of regular particles (dashed lines). Thus, energization in $E>B$ regions plays a key role in shaping the high-energy end of the spectrum.

\section{Conclusions} \label{conclusions}
We have employed 2D PIC simulations to quantify the impact of the energy gain occurring in $E>B$ (or equivalently, $\chi>0$) regions for the early stages of particle acceleration in relativistic reconnection with vanishing guide field. Most of our results are based on the calculation of the mean fractional contribution by $E>B$ fields (see $\zetaall$ in Eq.~\ref{eq:e_avg}) to particle energization up to the injection threshold $\estar$. We find that, in our simulations with outflow boundary conditions, the function $\zetaall$ is remarkably steady over time, as long as the reconnection rate has achieved a quasi-steady state ($\ttime\gtrsim 1.5$). At fixed threshold energy $\estar$, the fractional contribution $\zetaall$ monotonically increases with the particle energy $\et$ measured at time $T$. At fixed $\et$, we find that $\zetaall$ drops with increasing $\estar$. We show that $\zetaall$ increases with magnetization; for $\sigma\gtrsim 50$, we find that a fraction $\gtrsim 80\%$ of the energy that $\et/\sigma\gtrsim 8$ particles gain before crossing the $\estar=\sigma/4$ injection threshold is acquired in regions of electric dominance. 
As regard to the dependence on box size at fixed magnetization, we find it convenient to normalize $\et$ to the maximum particle energy $\epsilon_{\rm max}\propto L_{\rm x}^{1/2}$ (this scaling is appropriate only in 2D, see; e.g., \citet{zhang_21,zhang_23} for 3D); with this choice, we find that the curves of $\zetaall$ nearly overlap, when varying box size at fixed magnetization. 

We find that the distribution of energy gains acquired in $E>B$ regions can be modeled as in Eq.~\ref{eq:fit}. We have also demonstrated that energization in $E>B$ regions plays an important role in shaping the high-energy
end of the particle spectrum: we have evolved a population of test-particles, which are initialized and evolved as regular particles, but such that their energy is artificially kept fixed while in $E>B$ regions. We find that the spectra
of these test-particles are much steeper than the ones
of regular particles.

Our results help assessing the role of electric dominance in relativistic magnetic reconnection with vanishing guide field. \citet{Totorica_2023} demonstrated that non-ideal fields (i.e., $\boldsymbol{E}\neq-(\boldsymbol{v}_{\rm fl}/c) \times \boldsymbol{B}$) are essential for particle injection in relativistic reconnection. Regions of electric dominance certainly host non-ideal fields (i.e., $E>B$ is sufficient for the presence of non-ideal fields), yet  \citet{Totorica_2023} showed that plenty of non-ideal regions do not satisfy the condition for electric dominance. Near the midplane of the reconnection layer, the locations where non-ideal fields greatly exceed ideal fields (by more than a factor of two) generally correspond to $E>B$ regions. Most importantly, we argue that a large---possibly, dominant---fraction of the energy gain provided by non-ideal fields in zero-guide-field reconnection  should be ascribed to $E>B$ fields (see  Appendix \ref{periodic}, where we compare to the periodic runs by \citet{Totorica_2023}).


We conclude with a few caveats and potential future directions. First, our results are based on 2D simulations; \citet{sironi_22} and \citet{chernoglazov_23} showed good agreement between 2D and 3D simulations as regard to the injection physics, yet a dedicated study of the role of $E>B$ regions in 3D is still lacking. Second, our simulations employ an electron-positron plasma. The impact of $E>B$ fields for particle injection in electron-ion and electron-positron-ion plasma is left for a future study. Finally, we comment on the extension of our conclusions to non-relativistic, low-beta reconnection. The importance of electric dominance diminishes as $v_{\rm A}$ decreases:
the reconnected magnetic field scales linearly with distance from an X-point, whereas the reconnection electric field is $\sim 0.1 v_{\rm A} B_0$, so the extent of the region of electric dominance decreases as $\propto v_{\rm A}$ towards the non-relativistic regime.

\section*{Acknowledgements}

We thank F. Guo and S. Totorica for discussions on this topic. S.G. acknowledges support from the Barry Goldwater Scholarship. L.S. acknowledges support from DoE Early Career Award DE-SC0023015, NASA ATP 80NSSC24K1238, NASA ATP 80NSSC24K1826, and NSF AST-2307202. This work was supported by a grant from the Simons Foundation (MP-SCMPS-00001470) and facilitated by the Multimessenger Plasma Physics Center (MPPC), grant PHY-2206609. This project made use of the following computational resources: NASA Pleiades supercomputer, Ginsburg and Terremoto HPC clusters at Columbia University.

\section*{Data Availability}

The data underlying this paper will be shared upon reasonable request to the authors. 



\bibliographystyle{mnras}
\bibliography{workEgtB,araa.bib} 




\appendix

\section{Numerical Convergence} \label{spatial_res}

\begin{figure*}
 \includegraphics[width=\textwidth]{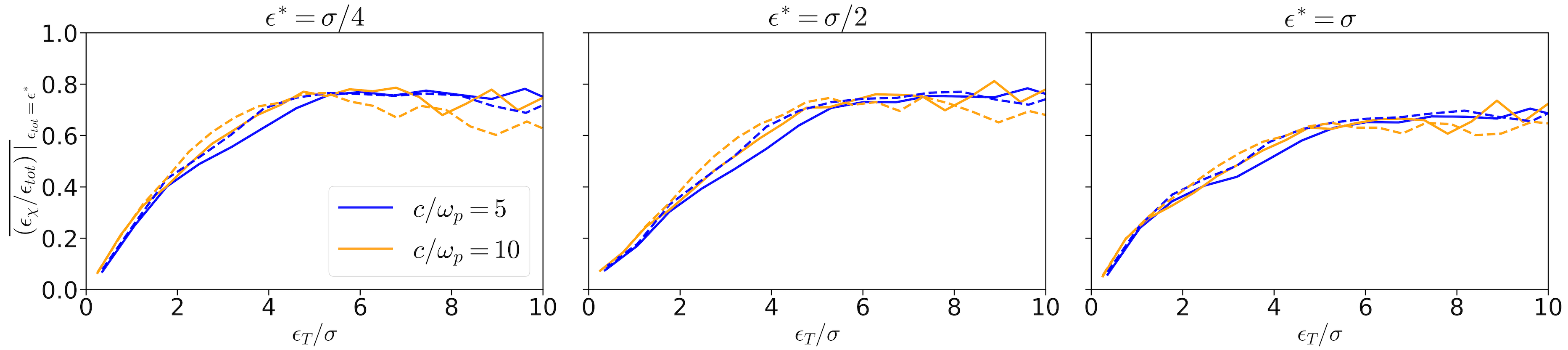}
 \caption{Same as Fig.~\ref{fig:time_evol}, but investigating the dependence on the spatial resolution, with $c/\omega_{\rm p} = 5$ (blue) and $10$ (orange). We use $\sigma=50$ and a box of $L_{\rm x}/(c/\omega_{\rm p}) = 384$ (so, smaller than our fiducial box). Solid lines are obtained by averaging in $2.5 < Tv_{\rm A}/L_{\rm x} < 3.1$, while dashed lines in $1.9 < Tv_{\rm A}/L_{\rm x} < 2.5$.
 }
 \label{fig:comp}
\end{figure*}

\begin{figure*}
 \includegraphics[width=\textwidth]{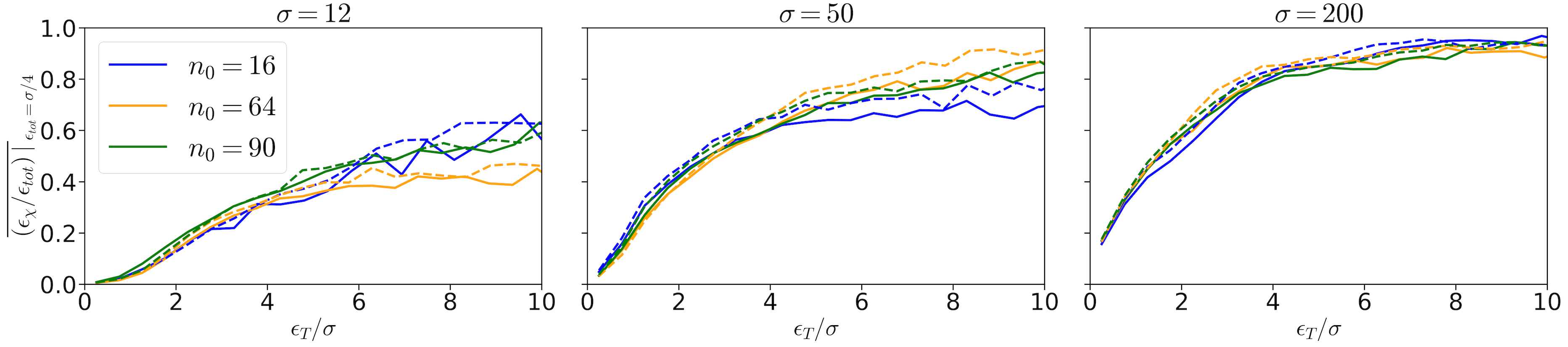}
 \caption{Same as Fig.~\ref{fig:time_evol}, but investigating the dependence on the number of particles per cell (see legend). We use $\estar=\sigma/4$ and the fiducial box of $L_{\rm x}/(c/\omega_{\rm p}) = 768$. Solid lines are obtained by averaging in $2.5 < Tv_{\rm A}/L_{\rm x} < 3.1$, while dashed lines in $1.9 < Tv_{\rm A}/L_{\rm x} < 2.5$.
 }
 \label{fig:sigma_ppc}
\end{figure*}

Fig.~\ref{fig:comp} shows the dependence of $\zetaall$
 on the spatial resolution, with $c/\omega_{\rm p} = 5$ (blue) and $10$ (orange). We use $\sigma=50$ and a box of $L_{\rm x}/(c/\omega_{\rm p}) = 384$ (so, smaller than our fiducial box). 
 Fig.~\ref{fig:sigma_ppc}, on the other hand, shows the dependence on the number of particles per cell (see legend). We use $\estar=\sigma/4$ and the fiducial box of $L_{\rm x}/(c/\omega_{\rm p}) = 768$. In both figures, solid lines are obtained by averaging in $2.5 < Tv_{\rm A}/L_{\rm x} < 3.1$, while dashed lines in $1.9 < Tv_{\rm A}/L_{\rm x} < 2.5$. Fig.~\ref{fig:comp} and Fig.~\ref{fig:sigma_ppc} shows that our results are numerically converged, both as regard to spatial resolution as well as number of particles per cell. 




\section{Simulations with Periodic Boundaries} \label{periodic}

\begin{figure*}
     \includegraphics[width=\textwidth]{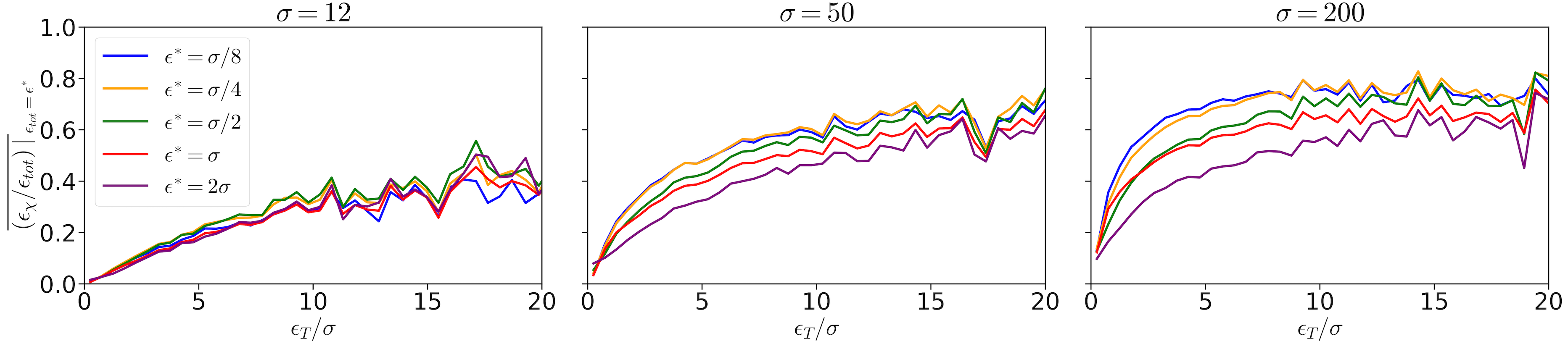}
    \caption{Same as Fig.~\ref{fig:time_evol}, but focusing on periodic boxes with $L_{\rm x}/(c/\omega_{\rm p}) = 768$. We vary the injection threshold (see legend) and the magnetization. We take our measurements at $Tv_{\rm A}/L_{\rm x}\sim4.0$. The horizontal axis extends up to $\epsilon_{\rm T}/\sigma=20$ to best compare with Figure 2(b) in \citet{Totorica_2023}.     
    }
    \label{fig:totorica_plot}
\end{figure*}

We compare our simulation results with those presented in Figure 2(b) of \citet{Totorica_2023}. They used periodic boundaries in $x$ and measured the fractional contribution of non-ideal fields at time $\omega_{\rm p}t\sim2000$, for a box with half-length $L_{\rm x}/(c/\omega_{\rm p})=500$; it follows that their measurements were taken at $\ttime\simeq 4.0$. Their fiducial run had $\sigma=50$ and $n_{0}=16$. We have run a set of three simulations (of varying $\sigma$) with periodic boundary conditions, including the $\sigma=50$ case considered by \citet{Totorica_2023}. Our box is larger (half-length $L_{\rm x}/(c/\omega_{\rm p})=768$) and our number of particles per cell is higher ($n_{0}=64$) than in \citet{Totorica_2023}, to retain some consistency with our outflow simulations.

The fractional contribution of $\chi>0$ regions to particle energization, as quantified by $\zetaall$ at $\ttime\simeq 4.0$, is shown in Fig.~\ref{fig:totorica_plot}. For each magnetization, we consider five threshold energies, $\epsilon^\ast = \sigma/8,\sigma/4, \sigma/2, \sigma, 2\sigma$ (shown in blue, orange, green, red, and purple, respectively), the same values used by \citet{Totorica_2023}. Unlike our earlier plots, the horizontal axis extends up to $\et/\sigma=20$ to best compare with Figure 2(b) in \citet{Totorica_2023}. In the following, we shall call $\zeta$ the fractional contribution of $E>B$ regions (as in the rest of our paper), while $\zeta_{\rm N}$ identifies the fractional contribution of non-ideal regions, as measured by \citet{Totorica_2023}. 
For $\sigma=50$---the case that \citet{Totorica_2023} showed in their Figure 2(b)---we find that for $\epsilon^\ast = \sigma/8$ and $\sigma/4$, $\zeta\simeq 0.8\, \zeta_{\rm N}$ at $\et/\sigma\simeq 5$, i.e., $\simeq 80\%$ of the non-ideal work is obtained in $E>B$ regions. The contribution is even higher at $\et/\sigma\gtrsim 10$, where $\zeta\simeq 0.85\, \zeta_{\rm N}$ for $ \epsilon^\ast = \sigma/8$ or $\sigma/4$. For higher threshold energies, $\epsilon^\ast = \sigma$ and $2\,\sigma$, we find that the fractional contribution of $E>B$ regions is greater than the one of non-ideal regions, $\zeta>\zeta_{\rm N}$ for all presented particle energies; our interpretation is that the non-ideal fields (with $E<B$) that particles encounter after their interaction with $E>B$ sites lead to deceleration, rather than further energization.

 


\section{Dependence on the Particle Age} \label{age}

\begin{figure*}
     \includegraphics[width=\textwidth]{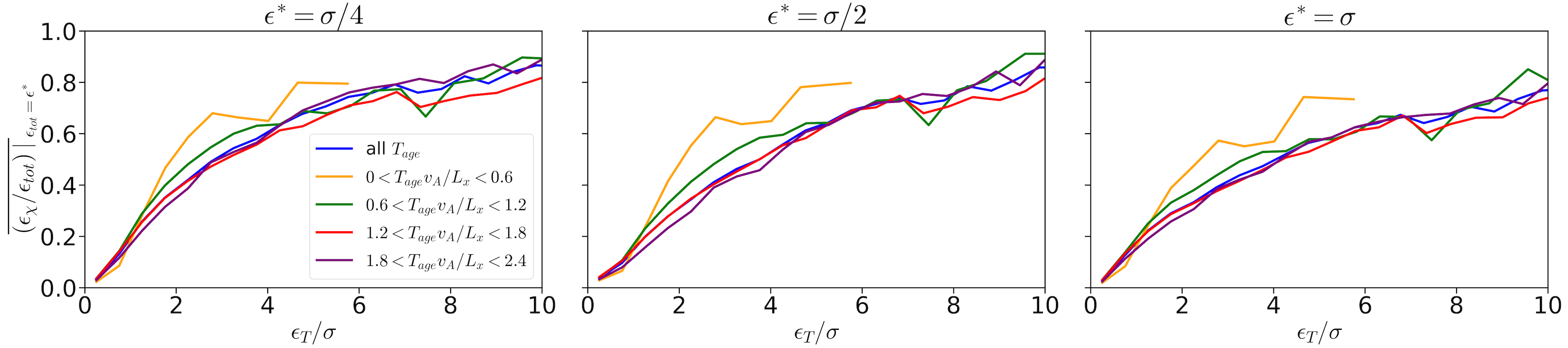}
    \caption{Same as Fig.~\ref{fig:time_evol}, but differentiating on the basis of the particle age, as indicated in the legend. The blue line includes all particles, regardless of their age (as we have done elsewhere). We consider our reference case with $\sigma=50$ and $L_{\rm x}/(c/\omega_{\rm p}) = 768$. All curves are time-averaged in the interval $2.5 < Tv_{\rm A}/L_{\rm x} < 3.1$.
    }
    \label{fig:particle_age}
\end{figure*}

Our calculation of $\zeta(\estar,\et)$ does not differentiate on the basis of the particle lifetime (i.e., how long a given particle needed to reach $\estar$ or $\et$, following its first interaction with the layer). In Fig.~\ref{fig:particle_age}, we show that  $\zeta(\estar,\et)$ is the same for particles having a wide range of lifetimes. We consider our reference case with $\sigma=50$ and $L_{\rm x}/(c/\omega_{\rm p}) = 768$. All curves are time-averaged in the interval $2.5 < Tv_{\rm A}/L_{\rm x} < 3.1$. We consider four different  age ranges, $0 < T_{\rm age}v_{\rm A}/L_{\rm x} < 0.6$, $0.6 < T_{\rm age}v_{\rm A}/L_{\rm x} < 1.2$, $1.2 < T_{\rm age}v_{\rm A}/L_{\rm x} < 1.8$, $1.8 < T_{\rm age}v_{\rm A}/L_{\rm x} < 2.4$ (shown in orange, green, red, and purple, respectively), and we compare them with the blue line, which includes all particles regardless of their age (as done elsewhere in the paper). We define birth  as the time when a particle first interacted with the layer (more precisely, the  time when its Lorentz factor first exceeds $\gamma=1.1$), and the particle age ($T_{\rm age}$) is the time interval from its birth to the time $T$ of our measurement. With the exception of the youngest particles with $0 < T_{\rm age}v_{\rm A}/L_{\rm x} < 0.6$, all curves overlap, suggesting that $\zetaall$ is insensitive to age.

\begin{figure*}
 \includegraphics[width=\textwidth]{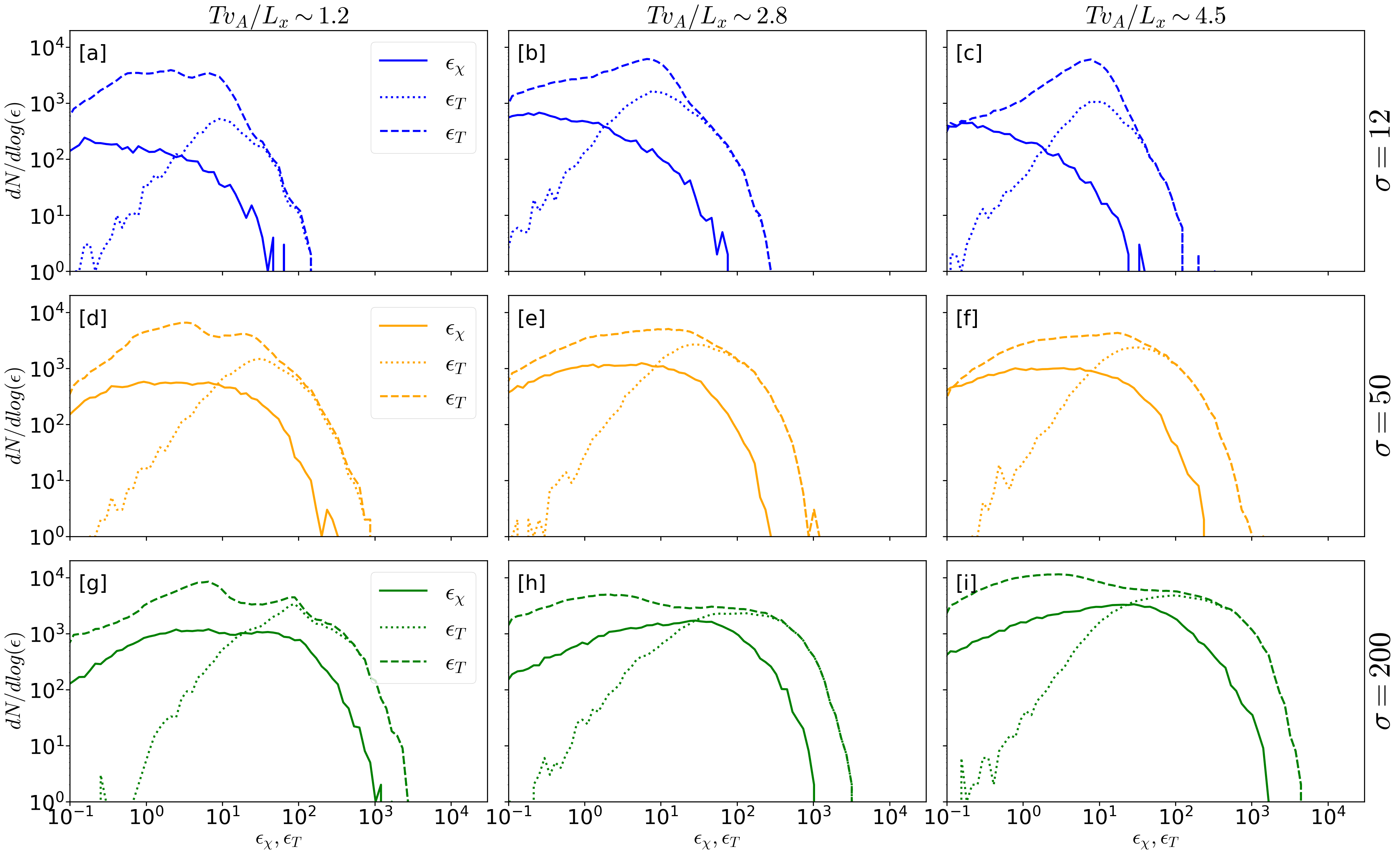}
 \caption{For different magnetizations (rows) and times $T$ (columns), we show: (a) the distribution of $\epsilon_{\rm \chi}=\gamma_\chi-1>0$, i.e., of the energy gains obtained in regions of electric dominance before time $T$ (solid lines); (b) the distribution of particle energies $\epsilon_{\rm T}$ for all the particles present at time $T$ (dashed lines); (c)  the distribution of particle energies $\epsilon_{\rm T}\mid_{\epsilon_{\rm \chi} > 0}$ only for the particles having $\epsilon_{\rm \chi}>0$ at time $T$ (dotted lines). 
}
 \label{fig:particle_counts}
\end{figure*}

\section{Time Dependence of the Energy Spectra} \label{time_energy}

\begin{table}
 \caption{Fraction of particles having experienced $E>B$}
 \begin{center}
 \begin{tabular}{ccc}
		\hline
        \hline
		\relax $\sigma$ & $Tv_{\rm A}/L_{\rm x}$ & $\epsilon_{\rm \chi} > 0$\\
		\hline
        \hline 
		12.5 & 1.2 & 0.07\\
		12.5 & 2.8 & 0.19\\
		12.5 & 4.5 & 0.16\\
        \hline
		50 & 1.2 & 0.14\\
        50 & 2.8 & 0.29\\
        50 & 4.5 & 0.31\\
        \hline
        200 & 1.2 & 0.22\\
		200 & 2.8 & 0.29\\
        200 & 4.5 & 0.27\\
	\end{tabular}
 \end{center}
 \label{tab:spectra_parameters}
 \begin{tablenotes}
 \small
 \item \relax \textit{Note}: For each time (as indicated in the second column), we report in the third column the fraction of particles that have experienced $E>B$ before that time.
 
 \end{tablenotes}
\end{table}

In Fig.~\ref{fig:particle_counts}, we extend the analysis presented in Fig.~\ref{fig:wpar_gam} to other times---$Tv_{\rm A}/L_{\rm x} = 1.2,2.8,4.5$.
For different magnetizations (rows) and times $T$ (columns), we show: (a) the distribution of $\epsilon_{\rm \chi}=\gamma_\chi-1>0$, i.e., of the energy gains obtained in regions of electric dominance before time $T$ (solid lines); (b) the distribution of particle energies $\epsilon_{\rm T}$ for all the particles present at time $T$ (dashed lines); (c)  the distribution of particle energies $\epsilon_{\rm T}\mid_{\epsilon_{\rm \chi} > 0}$ only for the particles having $\epsilon_{\rm \chi}>0$ at time $T$ (dotted lines). We complement this figure with Table \ref{tab:spectra_parameters}, where for each time $T$ we report the fraction of particles that have experienced $E>B$ before
that time. The main conclusion of Fig.~\ref{fig:particle_counts} is that at $\ttime\simeq 1.2$ the system has yet to achieve a steady state; in contrast, the results for $\ttime\simeq 2.8$ and 4.5 are generally consistent with each other.

\bsp	
\label{lastpage}
\end{document}